# Implementing distributed graph filters by elementary matrix decomposition

Samuel Cheng*, Senior Member, IEEE

*Abstract*—In this letter, we consider the implementation problem of distributed graph filters, where each node only has access to the signals of the current and its neighboring nodes. By using Gaussian elimination, we show that as long as the graph is connected, we can implement any graph filter by decomposing the filter into a product of directly implementable filters, filters that only use the signals at the current and neighboring nodes as inputs. We have also included a concrete example as an illustration.

## I. Introduction

Graph signal processing (GSP) extends classical digital signal processing (DSP) to signals on graphs, and provides potential solutions to numerous real-world problems that involve signals defined on topologically complicated domains, such as social networks, point clouds, biological networks, environmental and condition monitoring sensor networks [1], [2], [3], [4], [5]. However, there are several challenges in extending classical DSP to signals on graphs, particularly related to the scope of graph filters.

Despite many designs and applications of graph filters, efficient implementation of large graph filters has not been investigated much in the literature. In particular, for a "big data" graph that is stored in a distributed rather than a centralize manner [6]. In these scenarios, a graph node may only have access to graph signals physically near it, and in the extreme case, only to signals in its immediate neighborhood. To efficiently and distributively compute the result of a graph filter will be an interesting problem under this situation. The importance of this problem has also been reiterated in a recent GSP survey paper [1].

*Contribution:* in this letter, we focus on the completely distributed setup as mentioned previously (each node only has access to its own signal and the signals of its neighboring nodes). We introduce the concept of directly implementable filters and then we illustrate how using elementary matrix decomposition (Gaussian elimination) to decompose any filter into directly implementable filters and consequently facilitate a distributed implementation.

## II. Problem setup

By convention, an unweighted graph $G$ is defined by its vertex set $\mathcal{V}$ and its edge set $\mathcal{E}$. The vertex set contains all nodes (vertices) of the graph and the edge set contains the pair $(i, j)$ if and only if there is an edge between nodes $i$ and $j$. For an undirected graph, if $(i, j) \in \mathcal{E}$, then $(j, i) \in \mathcal{E}$ and vice versa.

We define a graph signal as a scalar number assigned to each vertex of the graph. We can vectorize that into a column vector $x$ of length $|\mathcal{V}|$. Similarly, a linear graph filter can be represented as a $|\mathcal{V}| \times |\mathcal{V}|$ matrix $M$, and the output of the graph filter is then simply the product $Mx$.

Under a distributed architecture, a node only has access to its own signal and signals of its immediate neighbors. Thus, a filter can be readily implemented only if the filter output at a target node only depends its own signal and its neighbors' signals. We call these directly implementable filters and define them precisely as follows.

*Definition 1 (Directly implementable filter).* Regarding to any undirected graph $G = (\mathcal{V}, \mathcal{E})$, we say a filter matrix $M$ of size $|\mathcal{V}| \times |\mathcal{V}|$ is directly implementable if $M_{j,i}$ can only be non-zero if $(i, j) \in \mathcal{E}$ or $i = j$. We denote $\mathcal{M}_G$ as the set of all directly implementable filter matrices regarding to the graph $G$.

*Example 1.* For the path graph on the right,

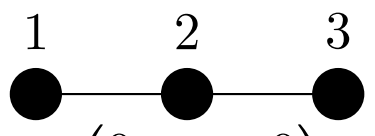

$\begin{pmatrix} 0 & a & 0 \\ b & 0 & c \\ 0 & d & 0 \end{pmatrix}$, $\begin{pmatrix} 0 & 0 & 0 \\ b & 0 & c \\ 0 & d & 0 \end{pmatrix}$ and $\begin{pmatrix} 1 & a & 0 \\ b & 0 & c \\ 0 & d & 0 \end{pmatrix}$ are directly implementable regarding to $G$ (i.e., belong to $\mathcal{M}_G$) but $\begin{pmatrix} 0 & a & 1 \\ b & 0 & c \\ 1 & d & 0 \end{pmatrix}$ is not.

*Remark 1.* Note that any diagonal matrix is directly implementable regarding to any undirected graph $G$.

How can we implement filters that are not directly implementable? Of course, we may send all signals into one centralized node and compute everything there. But it is really a centralized implementation and thus will not be considered here. Another more reasonable and obvious approach is to meticulously pass signals from each source node to the corresponding destination nodes. However, this would require nodes to also serve as routers for other nodes and significantly increase architectural complexity and communication cost. Consider the worst case scenario that each node need to facilitate traffic for all other nodes, the traffic cost alone will be of the order $O(n^3)$ for a graph with $n$ nodes.

S. Cheng is with the School of Electrical and Computer Engineering, University of Oklahoma, OK 74105, USA (email: samuel.cheng@ou.edu).

Instead, it would be much simpler if we can construct new filter by combining multiple directly implementable filters successively. The end result is a product filter. We denote the set of those product filters as below.

Definition 2 (Set of product filters). Denote $\mathcal{M}_G^{(k)} \triangleq \{M_k M_{k-1} \cdots M_1 | M_1, M_2, \cdots, M_k \in \mathcal{M}_G\}$

Note that the size of $\mathcal{M}_G^{(k)}$ grows as k increases and $\mathcal{M}_G^{(k)}$ encloses one another successively as stated below.

Remark 2. Note that since $I \in \mathcal{M}_G$, if $M \in \mathcal{M}_G$, $M = MI \in \mathcal{M}_G^{(2)}$. In general, $\mathcal{M}_G^{(k)} \subset \mathcal{M}_G^{(K)}$ for $k \leq K$.

## III. Main result

An interesting question to ask is what filters can be implemented as a product filter in $\mathcal{M}_G^{(k)}$. In particular, can we implement any filter with sufficiently large k? Note that each directly implementable filter only passes the signal of a target node to its immediately neighbors. Therefore, signal cannot flow more than k hops away for a filter in $\mathcal{M}_G^{(k)}$. So for a graph with diameter D, $\mathcal{M}_G^{(k)}$ apparently cannot cover all filters unless $k \geq D$. But how about $\mathcal{M}_G^{(D)}$? Will $\mathcal{M}_G^{(D)}$ be able to cover all possible filters?

The answer of the above question turns out to be negative. Consider a star graph with n nodes. Without loss of generality, say, Node 1 is connecting to all other nodes and there are no other edges. A directly implementable filter will have the form

$$\begin{pmatrix} m_{1,1} & m_{1,2} & \cdots \\ m_{2,1} & m_{22} & \\ \vdots & & \ddots \end{pmatrix},$$

where only the first row, the first column, and the diagonal can be non-zero. Thus the total degree of freedom is $3n-2$. Note that the diameter of any star graph is only 2. But for $\mathcal{M}_G^{(2)}$ to cover all possible filters, its degree of freedom must be at least $n^2$. Yet, the degree of freedom of $\mathcal{M}_G^{(2)}$ is only $2 \cdot (3n-2)$. So $\mathcal{M}_G^{(2)}$ cannot possibly cover all filter when $n > 5$.

On the other hand, since $|\mathcal{M}_G^{(k)}|$ increases as k increases, one may conjecture that for a sufficiently large k, $\mathcal{M}_G^{(k)}$ will be able to cover all filters. Indeed, we have the following theorem.

Theorem 1 (Graph filter decomposition). For any connected undirected graph, there exists a sufficiently large K such that any (filter) matrix M can be decomposed into the product $M_K M_{K-1} \cdots M_1$ with every matrix in the product being directly implementable (i.e., $M_1, M_2, \cdots, M_K \in \mathcal{M}_G$). In other words, $\mathcal{M}_G^{(K)}$ includes all possible filters.

Proof. Consider the elementary matrices defined as follows:

$$T_{i,j} = \begin{pmatrix} 1 & & & & & & \\ & \ddots & & & & & \\ & & 0 & & 1 & & \\ & & & \ddots & & & \\ & & 1 & & 0 & & \\ & & & & & \ddots & \\ & & & & & & 1 \end{pmatrix},$$

$$D_i(m) = \begin{pmatrix} 1 & & & & & & \\ & \ddots & & & & & \\ & & 1 & & & & \\ & & & m & & & \\ & & & & 1 & & \\ & & & & & \ddots & \\ & & & & & & 1 \end{pmatrix}, \text{ and}$$

$$L_{ij}(m) = \begin{pmatrix} 1 & & & & & & \\ & \ddots & & & & & \\ & & 1 & & & & \\ & & & \ddots & & & \\ & & m & & 1 & & \\ & & & & & \ddots & \\ & & & & & & 1 \end{pmatrix}.$$

Any matrices can then be diagonalized by the aforementioned elementary matrices via row reduction (Gaussian elimination) [7]. Namely, each column can be reduced to an almost all-zero vector (except the diagonal element potentially being non-zero) by applying $T_{i,j}$ (to swap in a non-zero diagonal term if the diagonal term is zero), $D_i(m)$ (to normalize the diagonal term to 1), followed by $n-1$ of $L_{i,j}(m)$ (to zero out the off-diagonal terms). And since the inverse of any elementary matrix is also elementary matrix (namely, $T_{i,j}^{-1} = T_{i,j}, D_i(m)^{-1} = D_i(1/m)$, and $L_{i,j}(m)^{-1} = L_{i,j}(-m)$), any matrix M can be decomposed into a product $M = E_n E_{n-1} \cdots E_1 D$, where D is a diagonal matrix and $E_1, E_2, \cdots, E_n$ are elementary matrices. Note that $D \in \mathcal{M}_G$ for any graph G and if we can show that any elementary matrix can be further decomposed into a product of matrices in $\mathcal{M}_G$, then the theorem is proved.

First, for $D_i(m)$, it is automatically in $\mathcal{M}_G$ since it is diagonal. For $T_{i,j}$, since G is connected, there is a path from i to j. Say the path is $i \to i_1 \to \cdots \to i_k \to j$, note that $T_{i,i_1}, T_{i_1,i_2}, \cdots, T_{i_{k-1},i_k}, T_{i_k,j}$ are all in $\mathcal{M}_G$. Moreover, it is easy to verify that

$$T_{i,j} = T_{i,i_1} T_{i_1,i_2} \cdots T_{i_{k-1},i_k} T_{i_k,j} T_{i_{k-1},i_k} \cdots T_{i_1,i_2} T_{i,i_1}, \quad (1)$$

so $T_{i,j}$ can be represented as a product of matrices in $\mathcal{M}_G$. Finally, for $L_{i,j}(m)$, just like the previous case, let say there exists a path from i to j as $i \to i_1 \to \cdots \to i_k \to j$.

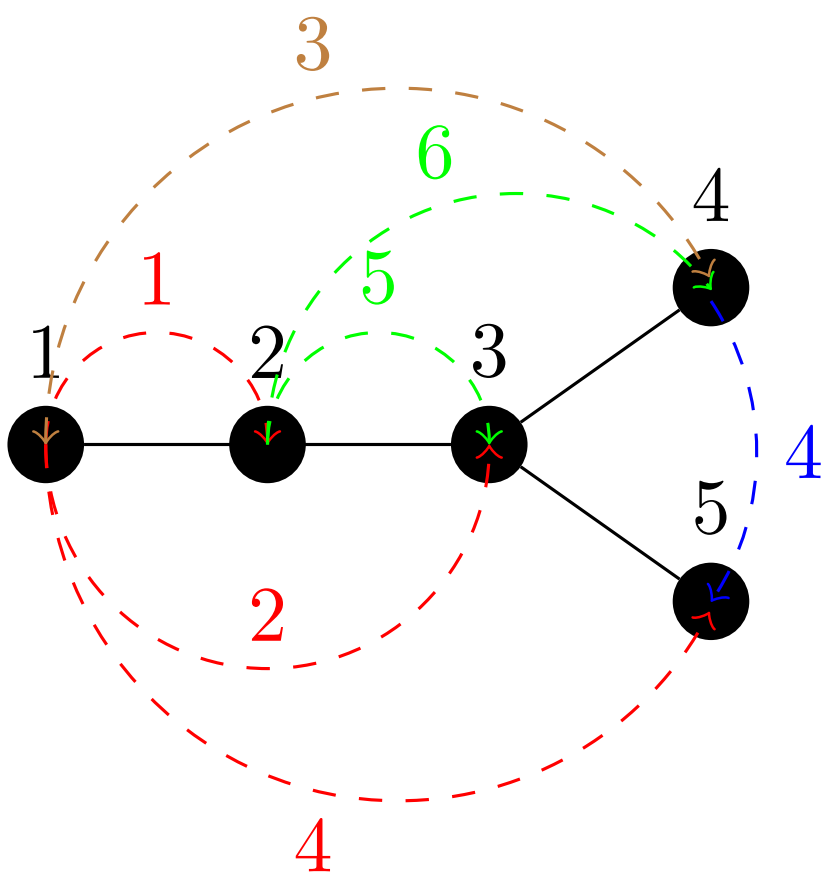

Fig. 1. Illustrating a graph filter applying on a 5-node graph (edges indicated by solid lines). The dash arrow lines indicate the "information flows" of the filter from starting to ending nodes. Note that the filter input does not always depend on its immediate neighbors (for instance, information flow is needed from Node 1 to Node 4) and thus the filter is not directly implementable under a distributed architecture.

Again, we have $T_{i,i_1}, T_{i_1,i_2}, \cdots, T_{i_{k-1},i_k}$ in $\mathcal{M}_G$ and so does $L_{i_k,j}(m)$. As before, we can easily verify that

$$L_{i,j}(m) = T_{i,i_1}T_{i_1,i_2}\cdots T_{i_{k-1},i_k}L_{i_k,j}(m)T_{i_{k-1},i_k}\cdots T_{i_1,i_2}T_{i,i_1}. \tag{2}$$

In summary, all elementary matrices can be represented as products of matrices in $\mathcal{M}_G$ and since any matrix can be decomposed into elementary matrices and a diagonal matrix, any matrices can be decomposed into matrices in $\mathcal{M}_G$.

Now, assume that the graph has a diameter $D$, which is the length of the longest path in $G$. Then, since $L_{i,j}(m)$ and $T_{i,j}$ can be decomposed into at most $2D-1$ directly implementable matrices, for the worst case scenario, any matrix can be represented by at most $n((2D-1)n+1)$ number of directly implementable filters. Since any $\mathcal{M}_G^{(k)}$ is a superset of $\mathcal{M}_G^{(K)}$ for $K \le k$ (cf. Remark 2), the proof is complete if we set $K = n((2D-1)n+1)$.

□

## IV. A simple example

Consider the 5-node graph as shown in Fig. 1 as an example. The solid lines in the figures represent the edges and the dash arrow lines indicate the information flows of the filter. Note that the filter input does not always depend on its immediate neighbors and thus not directly implementable (cf. Definition 1) under a distributed architecture.

As illustrated in Fig. 1, the filter can be represented in the matrix form as $\begin{pmatrix} 0&0&0&0&0\\ 1&0&0&3&0\\ 2&5&0&0&0\\ 3&6&0&0&0\\ 4&0&0&4&0 \end{pmatrix}$. By using Gaussian elimination, we can zero out the first column (except the first row) by multiplying several elementary matrices from the left as shown below:

$$L_{5,1}(-4)L_{4,1}(-3)L_{3,1}(-2)T_{1,2}\begin{pmatrix} 0&0&0&0&0\\ 1&0&0&3&0\\ 2&5&0&0&0\\ 3&6&0&0&0\\ 4&0&0&4&0 \end{pmatrix} =$$

$\begin{pmatrix} 1&0&0&3&0\\ 0&0&0&0&0\\ 0&5&0&-6&0\\ 0&6&0&-9&0\\ 0&0&0&-8&0 \end{pmatrix}$. Now, let's repeat the same procedure and zero out most of the second column (except the second row): $L_{4,2}(-6)D_2(1/5)T_{2,3}\begin{pmatrix} 1&0&0&3&0\\ 0&0&0&0&0\\ 0&5&0&-6&0\\ 0&6&0&-9&0\\ 0&0&0&-8&0 \end{pmatrix} =$

$\begin{pmatrix} 1&0&0&3&0\\ 0&1&0&-6/5&0\\ 0&0&0&0&0\\ 0&0&0&-9/5&0\\ 0&0&0&-8&0 \end{pmatrix}$. Finally, let's zero out the last (the fourth column) as follows:

$$L_{5,4}(8)L_{2,4}(6/5)L_{1,4}(-3)D_4(-5/9)\begin{pmatrix} 1&0&0&3&0\\ 0&1&0&-6/5&0\\ 0&0&0&0&0\\ 0&0&0&-9/5&0\\ 0&0&0&-8&0 \end{pmatrix}$$

$= \begin{pmatrix} 1&0&0&0&0\\ 0&1&0&0&0\\ 0&0&0&0&0\\ 0&0&0&1&0\\ 0&0&0&0&0 \end{pmatrix}$. Combining the equations above, the filter can then be decomposed into a product of elementary matrices as shown in (3), where the elementary matrices can be further decomposed into matrices in $\mathcal{M}_G$ (cf. (1) and (2) in the proof of Theorem 1) as follows:

$$\begin{aligned} L_{3,1}(2) &= T_{1,2}L_{3,2}(2)T_{1,2},\\ L_{4,1}(3) &= T_{1,2}T_{2,3}L_{4,3}(3)T_{2,3}T_{1,2},\\ L_{5,1}(4) &= T_{1,2}T_{2,3}L_{5,3}(4)T_{2,3}T_{1,2},\\ L_{4,2}(6) &= T_{2,3}L_{4,3}(6)T_{2,3},\\ L_{1,4}(3) &= T_{4,3}T_{3,2}L_{1,2}(3)T_{3,2}T_{4,3},\\ L_{2,4}(-6/5) &= T_{3,4}L_{2,3}(-6/5)T_{3,4},\\ L_{5,4}(-8) &= T_{3,4}L_{5,3}(-8)T_{3,4}. \end{aligned}$$

Substitute the above back into (3), we obtain (4). Note that several $T_{i,j}$ cancel each other in the intermediate step, and several elementary matrices can be grouped together (into square bracket terms) and yet are directly implementable. So the total number of directly implementable filters needed is at most 9[1].

## V. Conclusions and discussions

Through Gaussian elimination, we see that any graph filters can be decomposed into elementary matrices that

[1] There might be some other decompositions required fewer filters.

$$\mathrm{T}_{1,2}\mathrm{L}_{3,1}(2)\mathrm{L}_{4,1}(3)\mathrm{L}_{5,1}(4)\mathrm{T}_{2,3}\mathrm{D}_2(5)\mathrm{L}_{4,2}(6)\mathrm{D}_4(-9/5)\mathrm{L}_{1,4}(3)\mathrm{L}_{2,4}(-6/5)\mathrm{L}_{5,4}(-8)\begin{pmatrix}1&0&0&0&0\\0&1&0&0&0\\0&0&0&0&0\\0&0&0&1&0\\0&0&0&0&0\end{pmatrix}=\begin{pmatrix}0&0&0&0&0\\1&0&0&3&0\\2&5&0&0&0\\3&6&0&0&0\\4&0&0&4&0\end{pmatrix}, \tag{3}$$

$$\begin{aligned}\begin{pmatrix}0&0&0&0&0\\1&0&0&3&0\\2&5&0&0&0\\3&6&0&0&0\\4&0&0&4&0\end{pmatrix} &= \mathrm{T}_{1,2}(\mathrm{T}_{1,2}\mathrm{L}_{3,2}(2)\mathrm{T}_{1,2})(\mathrm{T}_{1,2}\mathrm{T}_{2,3}\mathrm{L}_{4,3}(3)\mathrm{T}_{2,3}\mathrm{T}_{1,2})(\mathrm{T}_{1,2}\mathrm{T}_{2,3}\mathrm{L}_{5,3}(4)\mathrm{T}_{2,3}\mathrm{T}_{1,2})\mathrm{T}_{2,3}\mathrm{D}_2(5)(\mathrm{T}_{2,3}\mathrm{L}_{4,3}(6)\mathrm{T}_{2,3})\\ &\quad \mathrm{D}_4(-9/5)(\mathrm{T}_{4,3}\mathrm{T}_{3,2}\mathrm{L}_{1,2}(3)\mathrm{T}_{3,2}\mathrm{T}_{4,3})(\mathrm{T}_{3,4}\mathrm{L}_{2,3}(-6/5)\mathrm{T}_{3,4})(\mathrm{T}_{3,4}\mathrm{L}_{5,3}(-8)\mathrm{T}_{3,4})\begin{pmatrix}1&0&0&0&0\\0&1&0&0&0\\0&0&0&0&0\\0&0&0&1&0\\0&0&0&0&0\end{pmatrix}\\ &= [\mathrm{L}_{3,2}(2)\mathrm{T}_{2,3}\mathrm{L}_{4,3}(3)\mathrm{L}_{5,3}(4)]\mathrm{T}_{2,3}[\mathrm{T}_{1,2}\underbrace{\mathrm{T}_{2,3}\mathrm{D}_2(5)\mathrm{T}_{2,3}}_{\mathrm{D}_3(5)}\mathrm{L}_{4,3}(6)]\mathrm{T}_{2,3}[\mathrm{D}_4(-9/5)\mathrm{T}_{4,3}][\mathrm{T}_{3,2}\mathrm{L}_{1,2}(3)]\\ &\quad [\mathrm{T}_{3,2}\mathrm{L}_{2,3}(-6/5)]\mathrm{L}_{5,3}(-8)\left[\mathrm{T}_{3,4}\begin{pmatrix}1&0&0&0&0\\0&1&0&0&0\\0&0&0&0&0\\0&0&0&1&0\\0&0&0&0&0\end{pmatrix}\right]\end{aligned} \tag{4}$$

can be further decomposed into directly implementable filters.

We see in the proof of Theorem 1 that for the worst case scenario, any matrix can be represented by at most $n((2D-1)n+1)$ number of directly implementable filters. However, this worst case bound is very loose in most cases as we can see from our numerical example that we can decompose the filter into 9 directly implementable filters rather than $5\cdot((2\cdot4-1)\cdot5-1)=170$ as given by the bound. As another example, consider an n-node fully connected graph and in that case any filter is directly implementable and so any filter can be decomposed into one directly implementable filter rather than $n(n+1)$ as given by the bound.

Note that we assumed undirected graphs in our discussion for the sake of clarity and easy exposition of the core idea. However, the result can be trivially extended to directed graphs. Basically, any graph filters can be decomposed into a product of directly implementable filters (regarding to a directed graph) in the similar manner as long as there is a path from any node to any other node.

A closely related line of research is to decompose a filter into a polynomial of "shift operations", where typically adjacency matrix or Laplacian matrix [8] will be used as the shift operator and the shift is, of course, directly implementable in general. However, only shift-invariant filters (filters commute with the shift operator) can be represented as such polynomials [9]. Moreover, a shift-enabled condition has to be satisfied by the shift operator itself [10], [11]. The solution considered in this letter is far more general as it applies to all possible linear filters.

## Acknowledgment

The author would like to thank Rick Ma and Charles Li for helpful discussions.